\documentclass[conference]{IEEEtran}
\IEEEoverridecommandlockouts
\usepackage{amsmath,amssymb,amsfonts}
\usepackage{algorithmic}
\usepackage{algorithm}
\usepackage{array}
\usepackage{textcomp}
\usepackage{stfloats}
\usepackage{bm}
\usepackage{url}
\usepackage{booktabs}
\usepackage{multirow}
\usepackage{verbatim}
\usepackage{graphicx}
\usepackage{cite}
\usepackage{hyperref}
\usepackage{amssymb}
\usepackage{cuted}
\usepackage[capitalise, nameinlink]{cleveref}
\usepackage{xcolor}
\usepackage{pstricks}
\usepackage{subfigure}

\def\BibTeX{{\rm B\kern-.05em{\sc i\kern-.025em b}\kern-.08em
    T\kern-.1667em\lower.7ex\hbox{E}\kern-.125emX}}
\begin{document}

\title{Multi-user ISAC through Stacked Intelligent Metasurfaces: New Algorithms and Experiments\\
\thanks{This work was supported by the National Natural Science Foundation of China under Grant 12141107.}
}

% \author{\IEEEauthorblockN{Ziqing~Wang}
% \IEEEauthorblockA{\textit{School of Electronic Information and Communications} \\
% \textit{Huazhong University of Science and Technology}\\
% Wuhan, China \\
% \and
% \and
% \and
% \and
% \and

\author{\IEEEauthorblockN{Ziqing Wang\textsuperscript{1}, Hongzheng Liu\textsuperscript{1}, Jianan Zhang\textsuperscript{1}, Rujing Xiong\textsuperscript{1}, Kai Wan\textsuperscript{1}, Xuewen~Qian\textsuperscript{2}, \\ Marco~Di~Renzo\textsuperscript{2},~Robert Caiming Qiu\textsuperscript{1}}
\IEEEauthorblockA{\textsuperscript{1} Huazhong University of Science and Technology, Wuhan, China. \\(e-mail: \{wangziqing,~hongzhengliu,~zhangjn,~rujing,~kai\_wan,~caiming\}@hust.edu.cn).\\
\textsuperscript{2} Universit{\'e} Paris-Saclay, CNRS, CentraleSup{\'e}lec, 
Laboratoire des Signaux et Systemes, 91192 Gif-sur-Yvette, France \\
(e-mail: xuewen.qian@centralesupelec.fr, marco.di-renzo@universite-paris-saclay.fr).}
}
% \author{\IEEEauthorblockN{Ziqing Wang, Hongzheng Liu, Jianan Zhang, Rujing Xiong, Kai Wan, Robert Caiming Qiu}
% \IEEEauthorblockA{Huazhong University of Science and Technology, Wuhan, China. \\(e-mail: \{wangziqing,~hongzhengliu,~zhangjn,~rujing,~kai\_wan,~caiming\}@hust.edu.cn).}
% }

\maketitle

\begin{abstract}
This paper investigates a Stacked Intelligent Metasurfaces (SIM)-assisted Integrated Sensing and Communications (ISAC) system. 
% The system comprises a multi-antenna base station (BS), a SIM formed by multiple transmissive reconfigurable intelligent surface (RIS) layers, multiple single-antenna communications users (CUs), and a non-line-of-sight (NLoS) target located within the BS's area. 
% The SIM can significantly improve channel quality, thus facilitating information transmission and target sensing in the ISAC system. 
An extended target model is considered, where the BS aims to estimate the complete target response matrix relative to the SIM. 
Under the constraints of minimum Signal-to-Interference-plus-Noise Ratio (SINR) for the communication users (CUs) and maximum transmit power, we jointly optimize the transmit beamforming at the base station (BS)  and the end-to-end transmission matrix of the SIM, to minimize the Cramér-Rao Bound (CRB) for target estimation. 
Effective algorithms such as the alternating optimization (AO) and semidefinite relaxation (SDR) are employed to solve the non-convex SINR-constrained CRB minimization problem. 
Finally, we design and build an experimental platform for SIM, and evaluate the performance of the proposed algorithms for communication and sensing tasks.
\end{abstract}


\section{Introduction}
Integrated Sensing and Communications (ISAC) has garnered increasing attention recently, driven by the forthcoming deployment of sixth-generation (6G) and subsequent communication systems\cite{liu2022survey}. 
With the expansion of frequency bands for 6G networks, 
% there arises the possibility of overlap between the frequency bands allocated to communications and radar systems. This overlap poses a risk of mutual interference \cite{liu2020joint}.
ISAC entails the fusion of sensing and communication systems to optimize congested wireless/hardware resources. 
This integrated approach embraces a symbiotic design paradigm, offering a unified platform for concurrently executing communication and radar sensing functionalities through the transmission of integrated signals \cite{liu2022integrated}.

In general, ISAC offers substantial improvements in performance and resource efficiency compared to standalone sensing and communication systems, largely due to the cooperative utilization of wireless resources, radio waveforms, and hardware platforms. However, sensing and communication functions, based on separate information processing principles, lead to various performance tradeoffs between them. 
These tradeoffs span from information-theoretic limits to physical layer performance considerations, as well as cross-layer design compromises. 
More precisely, from an information-theoretic viewpoint, the fundamental tradeoff between communication and sensing in ISAC can be mainly classified into two types:
(I) the tradeoff between the subspaces occupied by communication and sensing signals during transmission (ST). 
The ST tradeoff could be quantified by power allocation approaches across orthogonal or quasi-orthogonal dimensions ~\cite{hua2023mimo, liu2021cramer, 2019MU}. 
More essentially, the ST tradeoff could also be quantified by assessing the disparity between the span of the ``communication subspace'' and  ``sensing subspace''~\cite{ahmadipour2022information}, where the eigenspaces of these channels do not align or when the water-filling strategies do not concentrate power on the dominant eigenvectors. 
(II) The tradeoff between the determinism and randomness of the signals transmitted for communication and sensing (DRT). 
A larger communication capacity requires more randomness; conversely, higher sensing performance requires less randomness.
Therefore, the essence of ISAC lies in the tradeoff between the determinism and randomness of signals \cite{lu2023random}.

While ISAC effectively mitigates spectrum scarcity, obstacles such as buildings or trees in outdoor settings, as well as furniture or walls in indoor environments, can obstruct signal propagation. 
Reconfigurable Intelligent Surface (RIS) is regarded as a promising solution for enhancing ISAC systems, because of its unique capability to steer wireless signals to desired destinations without necessitating additional energy or spectrum resources.
Different types of RISs have been studied, including reflecting, refracting, and hybrid implementations \cite{li2016transmission,zhang2020beyond}.
A reflective RIS redirects incident signals toward users located on the same side of the base station (BS). 
In contrast, signals can pass through a refracting (or transmissive) RIS toward users positioned on the opposite side of the BS. 
As for the hybrid type, the RIS functions dually, facilitating both reflection and refraction functionalities. 
%Regarding 
A hybrid RIS is also referred to as simultaneously reflecting and transmitting reconfigurable intelligent surfaces (STAR-RIS) or intelligent omni-surface (IOS). %it possesses dual functionalities, being capable of both reflection and transmission operations. 

As for reflective RIS-assisted ISAC systems, the authors of \cite{song2022joint} and \cite{song2023cramerrao} jointly optimized the beamforming of both the BS and RIS by using semidefinite relaxation (SDR) and alternating optimization techniques. 
Since the modeling of a single transmissive RIS and a single reflective RIS is mathematically identical under ideal modeling assumptions, studies on ISAC assisted by a single reflective RIS can be extrapolated to the corresponding transmissive RIS. 
STAR-RIS-assisted ISAC systems considered in~\cite{liu2023toward} indicated the effectiveness of joint beamforming and the superior gain of STAR-RIS over conventional reflective RIS for ISAC. 

Prior studies on RIS-assisted ISAC systems primarily relied on single-layer meta-surface configurations, which resulted in coverage blind spots~\cite{wu2019beamforming}.
%Leveraging the collaborative potential of multiple RISs to aid BS in addressing various LoS blocking scenarios involves deploying multiple RISs for independent single-hop reflection \cite{yang2021energy} or cooperative multi-hop reflection to enhance coverage areas \cite{huang2021multi}.
A recently proposed technology called stacked intelligent metasurface (SIM), which is comprised of multiple transmissive RIS units arranged in parallel, has been proven to be a potent method for performance improvement at a reduced implementation complexity~\cite{an2023stacked, an2023stacked1}. 
% \footnote{Multi-hop reflective RISs typically involve larger spatial requirements, potentially leading to significant energy leakage between RIS nodes.}
Since the multi-layer beamforming in SIM-aided systems enhances the spatial resolution of the primary beam, SIM has been employed to facilitate multi-user communication and Direction of Arrival (DoA) estimation. 
Nevertheless, there has been no research works exploring SIM-aided ISAC systems and experiments validating the performance of SIM by using hardware platforms.

\subsection{Contributions} 
%This paper considers  SIM-aided ISAC systems in Fig.~\ref{fig_SIM_Extend_target}. 
The main contributions of this paper are the following:

(1) \textbf{Algorithm:} A sub-wavelength SIM model \cite{an2023stacked} is employed in this paper. 
Due to the multi-layer structure of the SIM, coupled with the rank-$1$ constraint for the transmit power for communication and the signal-to-noise ratio constraint for sensing, the optimization problem becomes inherently non-convex. 
We propose a Multi-Layer Alternating Optimization (MAO) algorithm by using the Singular Value Decomposition (SVD) to relax the optimization problems and develop an efficient solution for them. 

(2) \textbf{Experiments:} We manufactured three $1$-bit transmissive RISs that consist of $16 \times 16$ unit cells to construct an SIM, as illustrated in Fig.~\ref{fig_SIM_struct}.
Also, we conducted experiments in a controlled darkroom environment to validate the performance of the proposed algorithm for SIM-assisted communication-only and sensing-only applications, respectively. 
% For SIM-assisted communications and sensing, the number of SIM layers was systematically increased from one to three, the impact of SIM on communications SINR was investigated by varying the inter-layer distance of SIM, and the performance of DoA estimation under SIM assistance was examined by altering the position of the transmitter (TX) and adjusting the inter-layer distance of the SIM.

\section{System model}
\begin{figure}[t]
\centering
\includegraphics[width=2.3in]{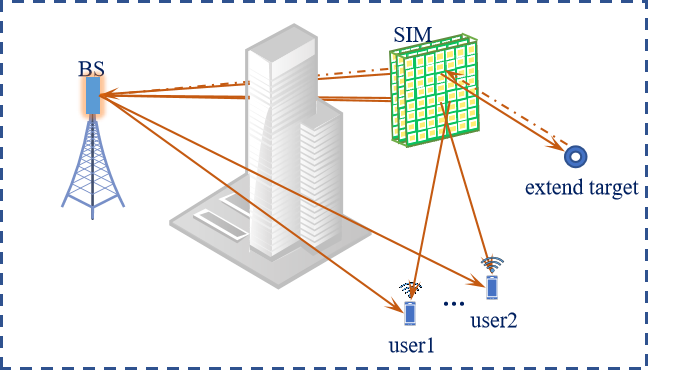}
\caption{System model of the considered SIM-aided multi-user ISAC system.} 
\label{fig_SIM_Extend_target}
\vspace{-0.4cm}
\end{figure}

In this paper, we investigate the SIM-aided ISAC system illustrated in Fig.~\ref{fig_SIM_Extend_target}. 
The formulation of the ISAC system is based on~\cite{song2023cramerrao}, with the main difference that we consider a SIM (i.e., multi-layer transmissive RIS) while~\cite{song2023cramerrao} considers a single-layer reflective RIS. 
The ISAC system consists of a BS with $M>1$ antennas, $K \ge 1$ single-antenna communication users (CUs), and an extended (sensing) target positioned within the Non-Line-of-Sight (NLoS) region of the BS. 
For the sake of simplicity, we assume that $K \le M$. 
The SIM is composed of $L$ metasurface layers, each of which consists of  $N$ meta-atoms  (unit cells).  
Let $\mathcal{L} = \left \{ 1,\dots, L \right \}$, $\mathcal{N} = \left \{ 1,\dots, N \right \}$, $\mathcal{K} = \left \{ 1,\dots, K\right \} $ denote the sets of metasurface layers, meta-atoms in each metasurface layer, and the set of users, respectively. 
According to Rayleigh-Sommerfeld diffraction theory \cite{lin2018all}, the transmission coefficient from the $\tilde{m}$-th meta-atom on the $(\ell- 1)$-th transmissive metasurface layer to the $m$-th meta-atom on the $\ell$-th transmissive layer is expressed by 
\begin{equation}
\omega_{m,\tilde{m}}^{\ell} = \frac{A_t\cos \chi_{m,\tilde{m}}^{\ell} }{r_{m,\tilde{m}}^{\ell}}\left ( \frac{1}{2\pi r_{m,\tilde{m}}^{\ell}}-j\frac{1}{\lambda } \right )e^{j2\pi r_{m,\tilde{m}}^{\ell}/\lambda }, \label{eq_trans_coeff}
\end{equation}
where $r_{m,\tilde{m}}^{\ell}$ denotes the transmission distance, $A_t$ denotes the area of the meta-atom, and $\chi_{m,\tilde{m}}^{\ell}$ represents the angle between the propagation direction and the normal direction of the ($\ell-1$)-th transmissive metasurface layer. 
Let $\mathbf{\Omega}_{\ell} \in \mathbb{C}^{N\times N}, \ell \in \mathcal{L}  $ denote the diffraction matrix between the $\ell$-th transmit metasurface layer and the ($\ell+1$)-th transmissive metasurface layer. 
$\mathbf{\Phi}_{\ell} \in \mathbb{C}^{N\times N}$ denotes the transmission coefficient matrix of the $\ell$-th transmissive metasurface layer, where $\bm{\Phi}_{\ell}=\operatorname{diag}(\bm{\phi}_{\ell,1}, \bm{\phi}_{\ell,2}, \ldots, \bm{\phi}_{\ell,N})$, $\bm{\phi}_{\ell,n} = e^{j\psi_{\ell,n}}$, with $\psi_{\ell,n} \in \left [ 0,2\pi \right )$ for each $n\in \mathcal{N}$. 
%$\left | \mathrm{\phi}_{\ell,n} \right | =1,\: \mathrm{arg}\left \{  \phi_{\ell,n} \right \} \in \left [ 0,2\pi \right )$. 
Therefore, the end-to-end transmission matrix of the SIM $\mathbf{P}\in \mathbb{C}^{N\times N}$ is expressed as
\begin{equation}
\mathbf{P} = \mathbf{\Phi}_{L}\mathbf{\Omega}_{L-1}\mathbf{\Phi}_{L-1} \dots \mathbf{\Phi}_{\ell}\mathbf{\Omega}_{\ell-1} \dots \mathbf{\Omega}_{2}\mathbf{\Phi}_{2}\mathbf{\Omega}_{1}\mathbf{\Phi}_{1}. \label{eq_trans_SIM}
\end{equation}

We consider a transmission block consisting of $T$ symbols. 
Let $\mathcal{T} = \left \{ 1,\dots,T \right \} $ denote the set of symbols. 
Also, let $\mathrm{s}_k(t)$ denote the communication signal for CU $k \in \mathcal{K}$ at time instant $t \in \mathcal{T}$ and $\mathbf{w}_k \in \mathbb{C}^{M \times 1}$ denote the transmission beamforming vector. 
Assume $\mathrm{s}_k(t)$ is an independent and identically distributed (i.i.d.) random variable with zero mean and unit variance. 
The transmitted signal at time instant $t$ is expressed as
\begin{equation}
\mathbf{x}(t) = \sum_{k \in \mathcal{K}} \mathbf{w}_k \mathrm{s}_k(t)+ \mathbf{x}_0, \label{signalforsac}
\end{equation}
where $\mathbf{x}_0 \in \mathbb{C}^{M \times 1}$ denotes the signal vector dedicated to sensing at symbol $t$, generated independently from $\mathrm{s}_k(t)$.

Let $\mathbf{h}^H_{d,k} \in \mathbb{C}^{1 \times N}$ denote the channel vector from the BS to CU $k$ for the direct link. 
Also, let $\mathbf{h}^H_{r,k} \in \mathbb{C}^{1 \times N}$ denote the channel vector from the SIM to CU $k$ and $\mathbf{G} \in \mathbb{C}^{N\times M}$ denote the channel matrix from the BS to the SIM. 
We assume that the BS has perfect channel state information (CSI) of the link between the BS and the CUs. 
This can be obtained through well-known channel estimation algorithms.
Then the received signal at CU $k$ and the time instant $t$ \cite{pan2022overview} is
\begin{equation}
\begin{aligned}
\mathrm{y}_k(t) &= ( \mathbf{h}^H_{d,k} + \mathbf{h}^H_{r,k} \mathbf{P} \mathbf{G}) \mathbf{x}(t) + \mathrm{n}_k(t) \\ &=  (\mathbf{h}^H_{d,k} + \mathbf{h}^H_{r,k} \mathbf{P} \mathbf{G}) \mathbf{w}_k \mathrm{s}_k(t) 
\\ & + (\mathbf{h}^H_{d,k} + \mathbf{h}^H_{r,k} \mathbf{P} \mathbf{G}) \sum_{i \in \mathcal{K}, i \neq k}\mathbf{w}_i \mathrm{s}_i(t) 
\\& + (\mathbf{h}^H_{d,k} + \mathbf{h}^H_{r,k} \mathbf{P} \mathbf{G}) \mathbf{x}_0 + \mathrm{n}_k(t), 
\label{eq_com_echo}
\end{aligned}
\end{equation}
where $\mathrm{n}_k(t) \sim \mathcal{CN} (0,\sigma_{k}^{2})$ denotes the additive white Gaussian noise (AWGN) at the CU $k$.

When transmitting $\mathbf{x}(t)$ for target sensing, the echoed signal matrix at the BS is formulated as
\begin{equation}
\begin{aligned}
\mathbf{y}_s(t) &= \mathbf{G}^T \mathbf{P}^T \mathbf{H}_s \mathbf{P} \mathbf{G} \mathbf{x}(t) + \mathbf{n}_R(t),\label{eq_extend_sensing_echo}
\end{aligned}
\end{equation}
where $\mathbf{H}_s\in \mathbb{C}^{N\times N}$ denotes the target response matrix (i.e., the cascaded channel over the SIM-target-SIM link) and $\mathbf{n}_R(t) \in \mathbb{C}^{M\times1}$ is the AWGN vector with variance $\sigma_R^2$.

For the extended target, 
% which may involve recognizing multiple objects, various regions, or multiple features, BS lacks prior knowledge of the quantity and associated angles of scatterers. 
% This is due to the random reflection of echoes in each of the BS's illuminations. 
based on~\cite{song2023cramerrao}, the CRB of the estimate of $\mathbf{H}_s$ is
\begin{small}
\begin{equation}
\begin{aligned}
&\mathbf{CRB}( \mathbf{H}_{s})\\& = \frac{\sigma_{\mathrm{R}}^2}{T} \operatorname{tr}\left (\left ( \mathbf{P}\mathbf{G} \mathbf{R}_{ie} \mathbf{G}^{H} \mathbf{P}^{H}  \right )^{-1}\right ) \operatorname{tr}\left ( \left ( \mathbf{P}\mathbf{G} \mathbf{G}^{H} \mathbf{P}^{H}  \right )^{-1}\right )\label{eqcrb1},
\end{aligned}
\end{equation}
\end{small}
where $\mathbf{R}_{ie} =  \sum_{k \in \mathcal{K}}  \mathbf{w}_k\mathbf{w}^{H}_k + \mathbf{R}_0$ with $\mathbf{R}_0 = \frac{1}{T}  \sum_{t \in \mathcal{T}} \mathbf{x}_0 \mathbf{x}_0^H$. 

Note that $\mathbf{H}_{s}$ can be estimated only when $\mathrm{rank}(\mathbf{P}\mathbf{G}) = N$ and $\mathrm{rank}(\mathbf{R}_{ie}) \ge N$, otherwise we have $\mathbf{CRB}(\mathbf{H}_s) \to \infty $. 
It is apparent that rank$(\sum_{k \in \mathcal{K}} \mathbf{w}_k \mathbf{w}_k^H) \le K$, therefore, when $K < N$, we need a dedicated sensing signal vector $\mathbf{x}_0$ to ensure rank$(\mathbf{R}_{ie}) \ge N$. 

Based on~\eqref{eq_com_echo}, the corresponding SINR of the $k$-th CU in the extended target case can be formulated as
\begin{small}
\begin{equation*}
\begin{aligned}
&\gamma_k=\\& \frac{\left|\mathbf{h}_{d,k}^{H} + \mathbf{h}_{r,k}^H \mathbf{P} \mathbf{G} \mathbf{w}_k\right|^2}{\sum_{i=1, i \neq k}^K\left|\mathbf{h}_{d,k}^{H} + \mathbf{h}_{r,k}^H \mathbf{P} \mathbf{G} \mathbf{w}_i\right|^2+\left\|\mathbf{h}_{d,k}^H+\mathbf{h}_{r,k}^H \mathbf{P} \mathbf{G} \mathbf{x}_0\right\|^2+\sigma_k^2}.
\end{aligned}
\end{equation*}
\end{small}

Let $\mathrm{P}_0$ be the maximum transmission power at the BS, the power constraint can be formulated as
\begin{equation}
\mathbb{E}(\left \| \mathbf{x}(t)\right \|^{2}  )  = \sum_{k \in \mathcal{K}} \left \| \mathbf{w}_k  \right \|^{2} + \operatorname{tr}(\mathbf{R}_0) \le \mathrm{P}_0. \label{eq_pow_con_extend}
\end{equation}

Then, we aim to minimize the $\mathbf{CRB}(\mathbf{H}_s)$ for the extended target estimation by jointly optimizing the transmission beamforming matrices $\left\{ \mathbf{w}_k \right\}$ and $\mathbf{R}_0$ at the BS, along with the end-to-end transmission matrix $\mathbf{P}$ of the SIM. 
The optimization problem needs to be solved while satisfying the minimum SINR constraints $\Gamma_k$ at the CUs and subject to the maximum transmit power constraint at the BS. 
Consequently, the problem can be formulated as $\mathrm{P}1$.
\begin{figure*}[!t]
\vspace{-0.3 cm}
\begin{subequations}
\begin{align}
(\mathrm{P}1): 
&\min _{\left\{\mathbf{w}_k\right\}, \mathbf{R}_0, \mathbf{P}} \operatorname{tr}((\mathbf{P G} \mathbf{R}_{ie} \mathbf{G}^H \mathbf{P}^H)^{-1}) \operatorname{tr}((\mathbf{P G G}^H \mathbf{P}^H)^{-1}) \notag \\
& \mathrm{s.t.} \quad \frac{\left| \left(\mathbf{h}_{\mathrm{d}, k}^H +\mathbf{h}_{\mathrm{r}, k}^H \mathbf{P G}\right) \mathbf{w}_k\right|^2}
{\sum_{i \in \mathcal{K}, i \neq k}
\left|  (\mathbf{h}_{\mathrm{d}, k}^H +\mathbf{h}_{\mathrm{t}, k}^H \mathbf{P G}) \mathbf{w}_i\right|^2+
(\mathbf{h}_{\mathrm{d}, k}^H +\mathbf{h}_{\mathrm{r}, k}^H \mathbf{P G}) \mathbf{R}_0 (\mathbf{h}_{\mathrm{d}, k}+ \mathbf{G}^H \mathbf{P}^H \mathbf{h}_{\mathbf{r}, k})
+\sigma_k^2} \geq \Gamma_k, \forall k \in \mathcal{K} \label{eq_extend_sinr}\\
& \sum_{k \in \mathcal{K}}\left\|\mathbf{w}_k\right\|^2+\operatorname{tr} (\mathbf{R}_0) \leq \mathrm{P}_0 \label{eq_extend_power}\\
 & \mathbf{R}_0 \succeq \mathbf{0} \label{eq_extend_positive}\\
 & \left|\mathbf{\phi}_{\ell,n}\right|=1, \forall \ell \in \mathcal{L}, \forall n \in \mathcal{N}.  \label{eq_rank_1_extend}
\end{align} 
\end{subequations}
\vspace{-1 cm}
\end{figure*}
It is worth noting that the SINR constraints in  \eqref{eq_extend_sinr} and the unit-modulus constraints in \eqref{eq_rank_1_extend} render problem $\mathrm{P}1$  non-convex. 
% Additionally, the interdependency between transmit and overall transmissive beamforming further complicates the non-convex nature of problem $\mathrm{P}1$. 
% Consequently, solving $\mathrm{P}1$ poses substantial challenges.

\section{SIM-aided Integrated Sensing and Communications}
In this section, we aim to solve problem $\mathrm{P}1$. 
Compared to the joint beamforming for ISAC with a single-layer reflective RIS in~\cite{song2023cramerrao}, the main challenge of the formulated problem resides in the multi-layer structure of the SIM. 
To address this challenge, we devise an efficient algorithm for problem $\mathrm{P}1$ by employing alternating optimization, to optimize  $\left\{\mathbf{w}_k \right\}$,  $\mathbf{R}_0$ and the end-to-end transmission matrix $\mathbf{P}$. 
\paragraph{Transmission Beamforming Optimization $\left\{\mathbf{w}_k \right\}$ and $\mathbf{R}_0$}
First, we optimize 
$\left\{\mathbf{w}_k \right\}$ and $\mathbf{R}_0$ in $\mathrm{P}1$ assuming  that
the end-to-end transmission matrix $\mathbf{P}$ of the SIM in~\eqref{eq_trans_SIM} is fixed  (i.e., assuming that $\mathbf{\Phi}_L,\ldots,\mathbf{\Phi}_1$ are fixed). 
So problem $\mathrm{P}1$ can be transformed as follows:
\begin{equation*}
\begin{aligned}
\mathbf(\mathrm{P}1.1): \min_{\{\mathbf{w}_k\}, \mathbf{R}_0} & \operatorname{tr} (  ( \mathbf{P}\mathbf{G}\mathbf{R}_{ie} \mathbf{G}^{H} \mathbf{P}^{H}  )^{-1} )\operatorname{tr}((\mathbf{P G G}^H \mathbf{P}^H)^{-1}) \nonumber  \\ \mbox{s.t.} \quad & \eqref{eq_extend_sinr},\eqref{eq_extend_power},\eqref{eq_extend_positive}.\nonumber
\end{aligned}
\end{equation*}

%The matrix $\mathbf{P}$ represents a multi-layer phase-shifting matrix of SIM, overall optimization is relatively complex. Based on the multi-layer structure of the SIM, 
% where each layer is independent, 
Based on the multi-layer structure of the SIM, we employ a layer-by-layer (i.e., optimizing $\mathbf{\Phi}_L,\ldots,\mathbf{\Phi}_1$ alternately) optimization approach for the matrices of transmission coefficient of the layers. 
The objective function in
problem P1.1 can be transformed as

% , keeping the others unchanged while optimizing the $L$-th layer $\mathbf{\Phi}_L$. 
The objective function in problem $\mathrm{P}1.1$ can be transformed as
\begin{small}
\begin{equation}
\begin{aligned}
&\operatorname{tr} (( \mathbf{P}\mathbf{G}\mathbf{R}_{ie} \mathbf{G}^{H} \mathbf{P}^{H})^{-1} ) \operatorname{tr}((\mathbf{P}\mathbf{G} \mathbf{G}^{H} \mathbf{P}^{H} )^{-1})
\\ =& \operatorname{tr} (  \underbrace{(\mathbf{\Phi}_L^H\mathbf{\Phi}_L)^{-1}}_{\mathbf{\Phi}_{L} \mathbf{\Phi}_{L}^{H} = \mathbf{I}_{N}}(\mathbf{A}_L\mathbf{G}\mathbf{R}_{ie} \mathbf{G}^{H} \mathbf{A}_L^{H})^{-1} ) \operatorname{tr}((\mathbf{P}\mathbf{G} \mathbf{G}^{H} \mathbf{P}^{H} )^{-1}) 
\\ =& \operatorname{tr} ( ( \mathbf{A}_L\mathbf{G}\mathbf{R}_{ie} \mathbf{G}^{H} \mathbf{A}_L^{H})^{-1} ) \operatorname{tr}((\mathbf{A}_L\mathbf{G} \mathbf{G}^{H} \mathbf{A}_L^{H} )^{-1}), \label{extend_multi_layer}
\end{aligned}
\end{equation}
\end{small}
where $\mathbf{A}_{\ell} := \mathbf{\Omega}_{\ell-1}\mathbf{\Phi}_{\ell-1} \dots \mathbf{\Omega}_{2}\mathbf{\Phi}_{2}\mathbf{\Omega}_{1}\mathbf{\Phi}_{1}$, for $\ell \in\{1,2,\ldots,L\}$. Hence,   problem $\mathrm{P}1.1$ can be further transformed as follows:
% \begin{equation*}
% \begin{aligned}
% \mathbf(\mathrm{P}1.2): \min_{\{\mathbf{w}_k\}, \mathbf{R}_0} \quad & \operatorname{tr}\left ( \left ( \mathbf{A}_L\mathbf{G}\mathbf{R}_{ie} \mathbf{G}^{H} \mathbf{A}_L^{H}  \right )^{-1}\right ) \nonumber  \\ \mbox{s.t.}\quad & \eqref{eq_extend_sinr},\eqref{eq_extend_power},\eqref{eq_extend_positive},\nonumber
% \end{aligned}
% \end{equation*}
\begin{small}
\begin{subequations}
\begin{align}
(\mathrm{P}1.2):&\min _{\left\{w_k\right\}, \mathbf{R}_0} \operatorname{tr} ((\mathbf{A}_L\mathbf{G}\mathbf{R}_{ie} \mathbf{G}^{H} \mathbf{A}_L^{H})^{-1}) \operatorname{tr}((\mathbf{A}_L\mathbf{G} \mathbf{G}^{H} \mathbf{A}_L^{H} )^{-1})\nonumber \\
&\mbox{s.t.} \quad (1+\frac{1}{\Gamma_k}) \operatorname{tr}(\mathbf{H}_k \tilde{\mathbf{W}}_k) -\operatorname{tr}( \mathbf{H}_k  (\sum_{k \in \mathcal{K}} \tilde{\mathbf{W}}_k+\mathbf{R}_0)) \geq \sigma_k^2,\notag 
\\& \ \ \ \ \ \  \forall k \in \mathcal{K}, \label{eq57}\\ 
&\sum_{k \in \mathcal{K}} \operatorname{tr}(\tilde{\mathbf{W}}_k)+\operatorname{tr}\left(\mathbf{R}_0\right) \leq \mathrm{P}_0, \\ 
&\mathbf{R}_0 \succeq \mathbf{0}, \tilde{\mathbf{W}}_k \succeq \mathbf{0}, \forall k \in \mathcal{K}, \\ &\operatorname{rank}(\tilde{\mathbf{W}}_k) \leq 1, \forall k \in \mathcal{K}, \label{eqrank-one}
\end{align}
\end{subequations}
\end{small}%
where $\mathbf{h}_k=\mathbf{h}_{\mathrm{d}, k}+\mathbf{G}^H \mathbf{A}_L^H \boldsymbol{\Phi}_{L}^{H} \mathbf{h}_{\mathbf{r}, k}$, $\tilde{\mathbf{W}}_k=\mathbf{w}_k\mathbf{w}_k^H$ and $ \mathbf{H}_k=\mathbf{h}_k \mathbf{h}_k^H$.
By relaxing the rank-$1$ constraint in~\eqref{eqrank-one}, problem $\mathrm{P}1.2$ becomes a convex semi-definite program (SDP), which can be solved by convex solvers such as CVX in~\cite{2008CVX}. 
The solutions for $\left\{\tilde{\mathbf{W}}_k^{\star} \right\}$ and ${\mathbf{R}_0^{\star}}$ by CVX for problem $\mathrm{P}1.2$ are 
\begin{subequations}
\begin{align}
\mathbf{w}_k^{\mathrm{opt}}=\left(\mathbf{h}_k^H \tilde{\mathbf{W}}_k^{\star} \mathbf{h}_k\right)^{-1 / 2} \tilde{\mathbf{W}}_k^{\star} \mathbf{h}_k, \forall k \in \mathcal{K}, \label{eq_w_extend_opt} \\ \mathbf{R}_0^{\mathrm{opt}}=\mathbf{R}_0^{\star}+\sum_{k \in \mathcal{K}} \tilde{\mathbf{W}}_k^{\star}-\sum_{k \in \mathcal{K}} \mathbf{w}_k^{\mathrm{opt}}\left(\mathbf{w}_k^{\mathrm{opt}}\right)^H.  \label{eq_R_opt} 
\end{align}
\end{subequations}
\paragraph{Transmission Coefficient Matrix $\mathbf{\Phi}_L$ Optimization}
We then optimize the end-to-end transmission matrix $\mathbf{P}$ in problem~$\mathrm{P}1$ with fixed $\left\{\mathbf{w}_k\right\}$ and $\mathbf{R}_0$.
The transmission coefficient matrices $\mathbf{\Phi}_1, \mathbf{\Phi}_2,\ldots,\mathbf{\Phi}_L$ in $\mathbf{P}$ will be optimized alternately.
% Next, the optimization focuses on the end-to-end transmission matrix $\mathbf{P}$ beginning with $\mathbf{\Phi}_L$ of the $L$-th layer of the SIM, considering the provided transmission beamforming matrices $\left\{\mathbf{w}_k\right\}$ and $\mathbf{R}_0$ in $\mathrm{P}1.3$. 

Since $\mathbf{CRB}(\mathbf{H}_s)$ in problem~$\mathrm{P}1.2$ is independent of the transmission coefficient matrix $\mathbf{\Phi}_L$,  we find $\mathbf{\Phi}_L$ with an explicit objective of enhancing the SINR at all CUs \cite{wu2019intelligent}. 
The optimization problem can be formulated as follows:
% \begin{equation*}
% \begin{aligned}
% (\mathrm{P}1.3): \mathbf{Find} \quad & \mathbf{\Phi}_L {\red {\rm REWRITE~THIS, THIS~IS~NOT~AN~OBJECTIVE}}\\
% \mbox{s.t.} \quad & \eqref{eq_extend_sinr},\eqref{eq_rank_1_extend}.
% \end{aligned}
% \end{equation*}
\begin{subequations}
\begin{align}
(\mathrm{P}1.3) & \max _{\mathbf{\Phi}_L,\left\{\varrho_k\right\}} \quad  \sum_{k \in \mathcal{K}} \varrho_k  \notag\\
\mbox{s.t.} \quad & \left|\left(\mathbf{h}_{\mathrm{d}, k}^H+\mathbf{h}_{\mathbf{r}, k}^H \mathbf{A}_L \boldsymbol{\Phi}_{L} \mathbf{G}\right) \mathbf{w}_k\right|^2 \notag\\
&-\Gamma_k \sum_{i \in \mathcal{K}, i \neq k}\left|\left(\mathbf{h}_{\mathrm{d}, k}^H+\mathbf{h}_{\mathrm{r}, k}^H \mathbf{A}_L \boldsymbol{\Phi}_{L} \mathbf{G}\right) \mathbf{w}_i\right|^2 \notag\\
&-\Gamma_k\left(\mathbf{h}_{\mathrm{d}, k}^H+\mathbf{h}_{\mathrm{r}, k}^H \mathbf{A}_L \boldsymbol{\Phi}_{L} \mathbf{G}\right) \mathbf{R}_0(\mathbf{h}_{\mathrm{d}, k} \notag\\
&+\mathbf{G}^H \boldsymbol{\Phi}_{L}^H \mathbf{A}_L^H  \mathbf{h}_{\mathbf{r}, k}) -\Gamma_k \sigma_k^2 \geq \varrho_k, \forall k \in \mathcal{K} \label{1_3a}\\
&\varrho_k \geq 0, \forall k \in \mathcal{K} \label{1_3b}\\
&\left|\boldsymbol{\phi}_{\ell, n}\right|=1, \forall \ell \in \mathcal{L}, \forall n \in \mathcal{N}. \label{1_3c}
\end{align}
\end{subequations}
The auxiliary variable $\varrho_k$ represents the difference between the actual SINR of CU $k$ and the threshold $\Gamma_k$. 
Problem $\mathrm{P}1.3$ can be solved by using SDR and Gaussian randomization \cite{wu2019intelligent}.
\paragraph{Transmission Coefficient Matrices $\mathbf{\Phi}_{L-1}, \ldots,\mathbf{\Phi}_{1}$ Optimization} 
After updating the transmission coefficient matrix of the $L$-th layer $\boldsymbol{\Phi}_L$, we optimize the matrix for $(L-1)$-th layer $\boldsymbol{\Phi}_{L-1}$ to the first layer $\boldsymbol{\Phi}_{1}$ alternately, employing the following procedure. 
We illustrate how to solve $\boldsymbol{\Phi}_{L-1}$ as an example and the remaining transmission coefficient matrices can be solved similarly. 
Initially, we write the objective function as 
\begin{align}
&\operatorname{tr}\big((\mathbf{P G} \mathbf{R}_{ie} \mathbf{G}^H \mathbf{P}^H)^{-1}\big) \operatorname{tr}\big((\mathbf{P G} \mathbf{G}^H \mathbf{P}^H)^{-1}\big) \nonumber \\
& =\operatorname{tr}\big(\mathbf{\Xi}_1 \mathbf{\Xi}_2 \big)
\operatorname{tr}\big(\mathbf{\Xi}_1 \mathbf{\Xi}_3\big), 
\label{eq:tr tr obj}
\end{align}
where
\begin{subequations}
\begin{align}
\mathbf{\Xi}_1=&(\mathbf{\Omega}_{L-1}^{H} \mathbf{\Omega}_{L-1})^{-1},\\
\mathbf{\Xi}_2=&(\mathbf{\Phi}_{L-1} \mathbf{\Omega}_{L-2} \ldots \mathbf{\Phi}_2 \mathbf{\Omega}_1 \mathbf{\Phi}_1 \mathbf{G} \mathbf{R}_{ie} \mathbf{G}^H \nonumber \\
&(\mathbf{\Phi}_{L-1} \mathbf{\Omega}_{L-2} \ldots \mathbf{\Phi}_2 \mathbf{\Omega}_1 \mathbf{\Phi}_1)^H)^{-1}, \\
\mathbf{\Xi}_3=&(\mathbf{\Phi}_{L-1} \mathbf{\Omega}_{L-2} \ldots \mathbf{\Phi}_2 \mathbf{\Omega}_1 \mathbf{\Phi}_1 \mathbf{G}  \mathbf{G}^H \nonumber \\
&(\mathbf{\Phi}_{L-1} \mathbf{\Omega}_{L-2} \ldots \mathbf{\Phi}_2 \mathbf{\Omega}_1 \mathbf{\Phi}_1)^H)^{-1}. 
%\notag\\
%\mathbf{\Xi}_3=&(\mathbf{\Omega}_{L-1}^{H} \mathbf{\Omega}_{L-1})^{-1},\\
%\mathbf{\Xi}_4=&(\mathbf{\Phi}_{L-1} \mathbf{\Omega}_{L-2} \ldots \mathbf{\Phi}_2 \mathbf{\Omega}_1 \mathbf{\Phi}_1 \mathbf{G} \mathbf{G}^H\\
%&(\mathbf{\Phi}_{L-1} \mathbf{\Omega}_{L-2} \ldots \mathbf{\Phi}_2 \mathbf{\Omega}_1 \mathbf{\Phi}_1)^H)^{-1}. \notag
\end{align}
\end{subequations}
Based on the properties of the diffraction matrix $\mathbf{\Omega}_{L-1}$, $\mathbf{\Xi}_1$ is a fixed full-rank Hermitian matrix.  
We denote the singular value decomposition $\mathbf{\Xi}_1 = \mathbf{U}\mathbf{\Sigma V}$ with the maximum and minimum singular values denoted by $\sigma_{max}$ and $\sigma_{min}$,  respectively, which are constant. 
%$\mathbf{\Sigma} = \operatorname{diag}(\sigma_{max}, \dots, \sigma_{min})$, 
Hence, we have  $\sigma_{min} \operatorname{tr}(\mathbf{\Xi}_2) \leq \operatorname{tr}(\mathbf{\Xi}_1 \mathbf{\Xi}_2) =\operatorname{tr}\left(\mathbf{\Xi}_2^{1 / 2} \mathbf{\Xi}_1 \mathbf{\Xi}_2^{1 / 2}\right) \leq \sigma_{max} \operatorname{tr}(\mathbf{\Xi}_2)$. Thus we can take the following relaxation on the objective function in~\eqref{eq:tr tr obj}
\begin{align}
& \operatorname{tr}(\mathbf{\Xi}_1\mathbf{\Xi}_2) \rightarrow \operatorname{tr}(\mathbf{\Xi}_2). \label{EF}%\\
%& \min \operatorname{tr}(\mathbf{\Xi}_3\mathbf{\Xi}_4) \rightarrow \min \operatorname{tr}(\mathbf{\Xi}_4). \label{CD}
\end{align}
Similarly, we can also take
the following relaxation on the objective function in~\eqref{eq:tr tr obj}
\begin{align}
& \operatorname{tr}(\mathbf{\Xi}_1\mathbf{\Xi}_3) \rightarrow \operatorname{tr}(\mathbf{\Xi}_3). \label{EF2}%\\
%& \min \operatorname{tr}(\mathbf{\Xi}_3\mathbf{\Xi}_4) \rightarrow \min \operatorname{tr}(\mathbf{\Xi}_4). \label{CD}
\end{align}

As a result, the relaxed objective function in~\eqref{eq:tr tr obj} becomes 
$ \operatorname{tr}(\mathbf{\Xi}_2)\operatorname{tr}(\mathbf{\Xi}_3)$, which is independent of $\mathbf{\Phi}_{L-1}$. Based on the related objective function: 
\begin{itemize}
    \item we first update $\left\{\mathbf{w}_k \right\}$ and $\mathbf{R}_0$ by
problem $\mathrm{P}1.2$, but with the updated objective function $\operatorname{tr}(\mathbf{\Xi}_1\mathbf{\Xi}_3) \rightarrow \operatorname{tr}(\mathbf{\Xi}_3)$ and with   $\mathbf{h}_k=\mathbf{h}_{\mathrm{d}, k}+\mathbf{G}^H \mathbf{A}_{L-1}^H \boldsymbol{\Phi}_{L-1}^{H} \mathbf{h}_{\mathbf{r}, k}$, $\tilde{\mathbf{W}}_k=\mathbf{w}_k\mathbf{w}_k^H$ and $ \mathbf{H}_k=\mathbf{h}_k \mathbf{h}_k^H$;
\item we then obtain $\mathbf{\Phi}_{L-1}$ by solving $\mathrm{P}1.3$, in which we replace $\mathbf{A}_{L}$ by $\mathbf{A}_{L-1}$  and $\mathbf{\Phi}_{L}$ by $\mathbf{\Phi}_{L-1}$.
\end{itemize}

\begin{figure}[ht]
\centering
\includegraphics[width=3.5in]{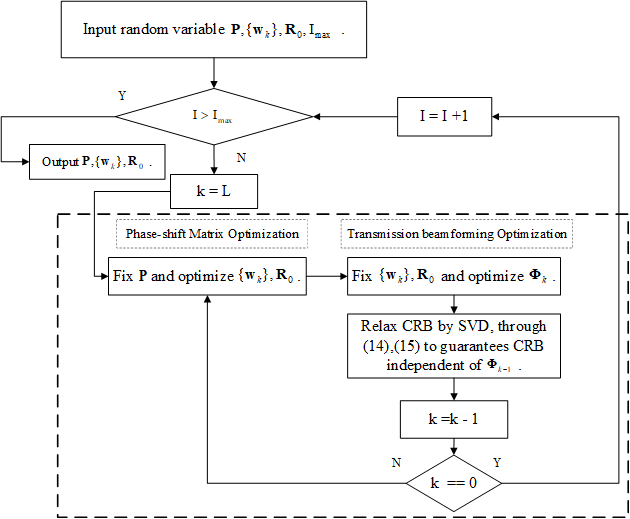}
\caption{Multi-layer Alternating Optimization algorithm (MAO). $I_{\rm max}$ represents the maximum number of iterations.} 
\label{fig_MAO}
\end{figure}

Similarly, the remaining phase shift matrices $\boldsymbol{\Phi}_{L-2},\ldots,  \boldsymbol{\Phi}_{1}$ are obtained by the same alternating methods.
A block diagram of the proposed algorithm is illustrated in Fig.~\ref{fig_MAO}. 
% Algorithm~\ref{alg:alg3}.

% The cost function for the transmission phase-shift matrix of the $(L-1)$-th layer of the SIM can be formulated as
% \begin{subequations}
% \begin{align}
% &\operatorname{tr}((\mathbf{\Phi}_{L-1} \mathbf{\Omega}_{L-2} \ldots \mathbf{\Phi}_2 \mathbf{\Omega}_1 \mathbf{\Phi}_1 \mathbf{G R}_{ie} \mathbf{G}^H \nonumber \\
% &(\mathbf{\Phi}_{L-1} \mathbf{\Omega}_{L-2} \ldots \mathbf{\Phi}_2 \mathbf{\Omega}_1 \mathbf{\Phi}_1)^H)^{-1}) \notag \\ 
% &=\operatorname{tr}((\mathbf{\Omega}_{L-2} \ldots \mathbf{\Phi}^2 \mathbf{\Omega}_1 \mathbf{\Phi}_1 \mathbf{G} \mathbf{R}_{ie} \mathbf{G}^H\nonumber\\
% &\left(\mathbf{\Omega}_{L-2} \ldots \mathbf{\Phi}_2 \mathbf{\Omega}_1 \mathbf{\Phi}_1\right)^H)^{-1}). \notag 
% \end{align}
% \end{subequations}

% The optimization problem for the phase-shift matrix of $(L-1)$th layer can be rewritten as $\mathrm{P}1.5$,

%similarly, the optimization problem on the phase-shift matrices for the $(L-1)$-th layer $\boldsymbol{\Phi}_{L-1}$ to the first layer $\boldsymbol{\Phi}_{1}$ can be also solved by the proposed algorithm summarized in Algorithm~\ref{alg:alg3}.
                     
\vspace{-0.4 cm}
\vspace{-0.3 cm}
\section{NUMERICAL RESULTS}
In this section, numerical results are introduced to evaluate the performance of our proposed algorithm. 
The distance-dependent path loss is set to $L(d) = L_0(\frac{d}{d_0})^{-\mathrm{\alpha}_0}$, where $L_0 = -30 \mathrm{dB}$ denotes the loss at the reference $d_0 = 1 {\rm m}$ and $\mathrm{\alpha}_0$ denotes the loss exponent. 
We adopt the Rician fading model for the BS-SIM and SIM-CUs links with the Rician factor equal to $0.5$. 
The coordinates of the BS and the SIM are $(0 {\rm m}, 0 {\rm m}, 11 {\rm m})$ and $(0 {\rm m}, 0 {\rm m}, 10 {\rm m})$, respectively, and two CUs are located at $(0 {\rm m}, 10 {\rm m}, 0 {\rm m})$, $(0 {\rm m}, 20 {\rm m}, 0 {\rm m})$, respectively.  
The SINR constraint for both CUs is set to $\Gamma \in [0 {\rm dB}, 30 {\rm dB}]$.
The extended (sensing) target is positioned within the
NLoS region of the BS. 
Other simulation parameters are defined as follows: $M = N = 4$, $P_0 = 120$ dBm, and $\sigma^{2}_R= -120$ dBm noise power. 
In addition, the thickness of the SIM is set to $T_{\mathrm{SIM}} = 3\lambda$ and the spacing between adjacent transmissive layers of the SIM with $L$ layers is $d_{\mathrm{SIM}} = T_{\mathrm{SIM}}/{(L-1)}$.

\begin{figure}[ht]
\vspace{-0.5 cm}
\centering
\includegraphics[width=2.1in]{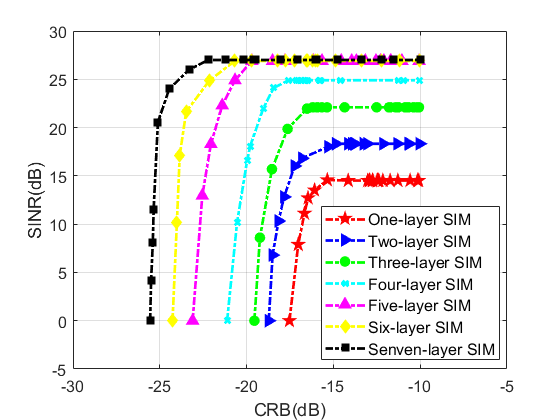}
\vspace{-0.2 cm}
\caption{The CRB for $\mathbf{H}_s$ estimation versus the SINR threshold.} 
\label{fig_CU_extend_ISAC}
\vspace{-0.2 cm}
\end{figure}

It can be observed in Fig.~\ref{fig_CU_extend_ISAC} that by increasing the number of layers in the SIM leads to an enhanced performance for ISAC.  
In the figure, specifically, we show the CRB for estimating ${\bf H}_s$ for different values of the SINR threshold. By increasing the number of layers, a gain of a dew decibel is obtained.

\section{Experimental Results}

% In this section, experimental validation was conducted to illustrate that the communications performance is enhanced with the increasing number of SIM layers, correlating with the spacing between the boards. 
% Additionally, the error in DoA estimation diminishes as the number of SIM layers increases, largely consistent with the theoretical results.
\subsection{The Experimental Platform}
\begin{figure}[ht]
\vspace{-0.5 cm}
\centering
\includegraphics[width = 3.2 in]{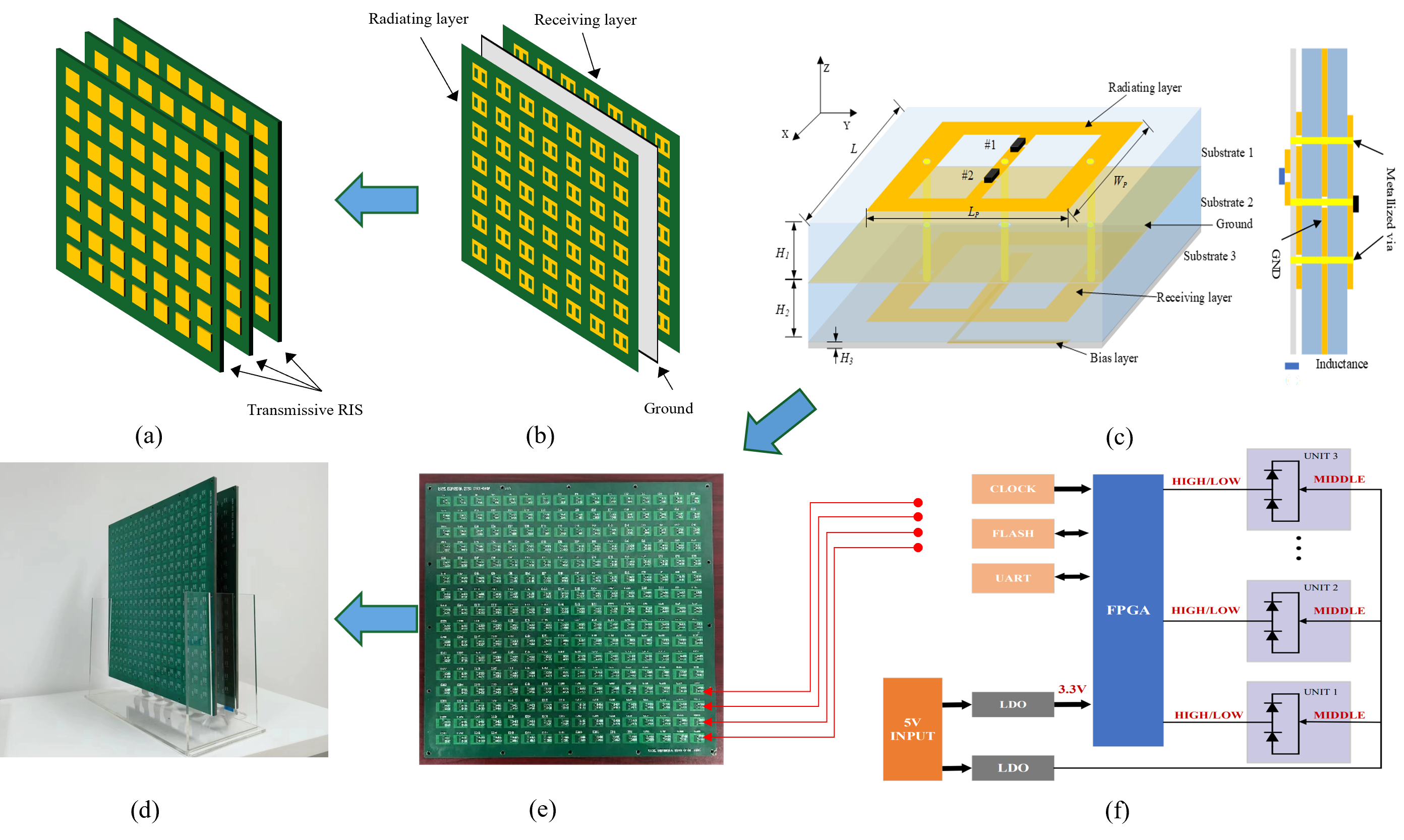}
\vspace{-0.5 cm}
\caption{Stacked Intelligent Metasurface (SIM). (a) The structure of SIM. (b) The structure of a transmissive layer. (c) The structure of a unit cell. (d) 1-bit SIM. (e) 1-bit transmissive layer. (f) Schematic diagram of the logic circuit board.} 
\label{fig_SIM_struct}
\vspace{-0.1 cm}
\end{figure}

In this section, we analyze the performance of SIM by utilizing a new hardware platform.
The SIM comprised of multiple transmissive metasurface layers and is shown in Fig.~\ref{fig_SIM_struct}a. 
Each meta-atom of the SIM can apply two phase shifts ($1$-bit or binary unit cell).

Each metasurface layer comprises three layers: the radiating layer, the receiving layer, and the ground plane, as depicted in Fig.~\ref{fig_SIM_struct}b. 
Additionally, since the influence of the transmissive plate on signals incident from both sides remains consistent, the structure of the radiating layer and the receiving layer is generally identical. 
We leverage printed circuit board (PCB) technology to simplify the fabrication process. 
It embraces the layout of a uniform planar array, comprising $16 \times 16$ unit cells, with total size $31 \times 31 \text{cm}^{2}$. %The transmissive RIS interfaced to eight connectors, each with $1 \times 34$ pins.
% Multiple $1$-bit transmissive RISs (as illustrated in Fig.~\ref{fig_SIM_struct}e) arranged in parallel can constitute a $1$-bit SIM (as illustrated in Fig.~\ref{fig_SIM_struct}d).

The geometry of each unit cell is illustrated in Fig.~\ref{fig_SIM_struct}c, consisting of four copper layers upheld by three substrate layers.
The copper layers, arranged from top to bottom, comprise the radiating layer, the ground plane, the receiving layer, and the bias layer \cite{zhang2024design}. 
% By dynamically modulating the quantized code distributions on the RIS, it becomes feasible to generate scanning beams.
The radiating layer comprises a rectangular annular patch connected to a rectangular patch, integrated with two PIN diodes (\#1 and \#2).
The outer rectangular annular patch is linked to the ground plane, whereas the rectangular inner patch is connected to the receiving layer via a metalized via-hole. 
Additionally, the receiving layer is interconnected with the bias layer via dielectric substrate 3 ($H_3 = 0.2 {\rm mm}$) through via-holes.
The substrate 1 and the substrate 2 layers are designed using F4B materials ($H_1 = H_2 = 2 {\rm mm}$).

The steering-logic board is designed and fabricated as in Fig.~\ref{fig_SIM_struct}f. 
It is powered by a voltage input supply ($5{\rm V}$), and an extra low dropout regulator (LDO) provides an external reference voltage for the ground plane (marked as MIDDLE). 
\vspace{-0.4cm}
\begin{figure}[ht]
\centering
\includegraphics[width=2.1in]{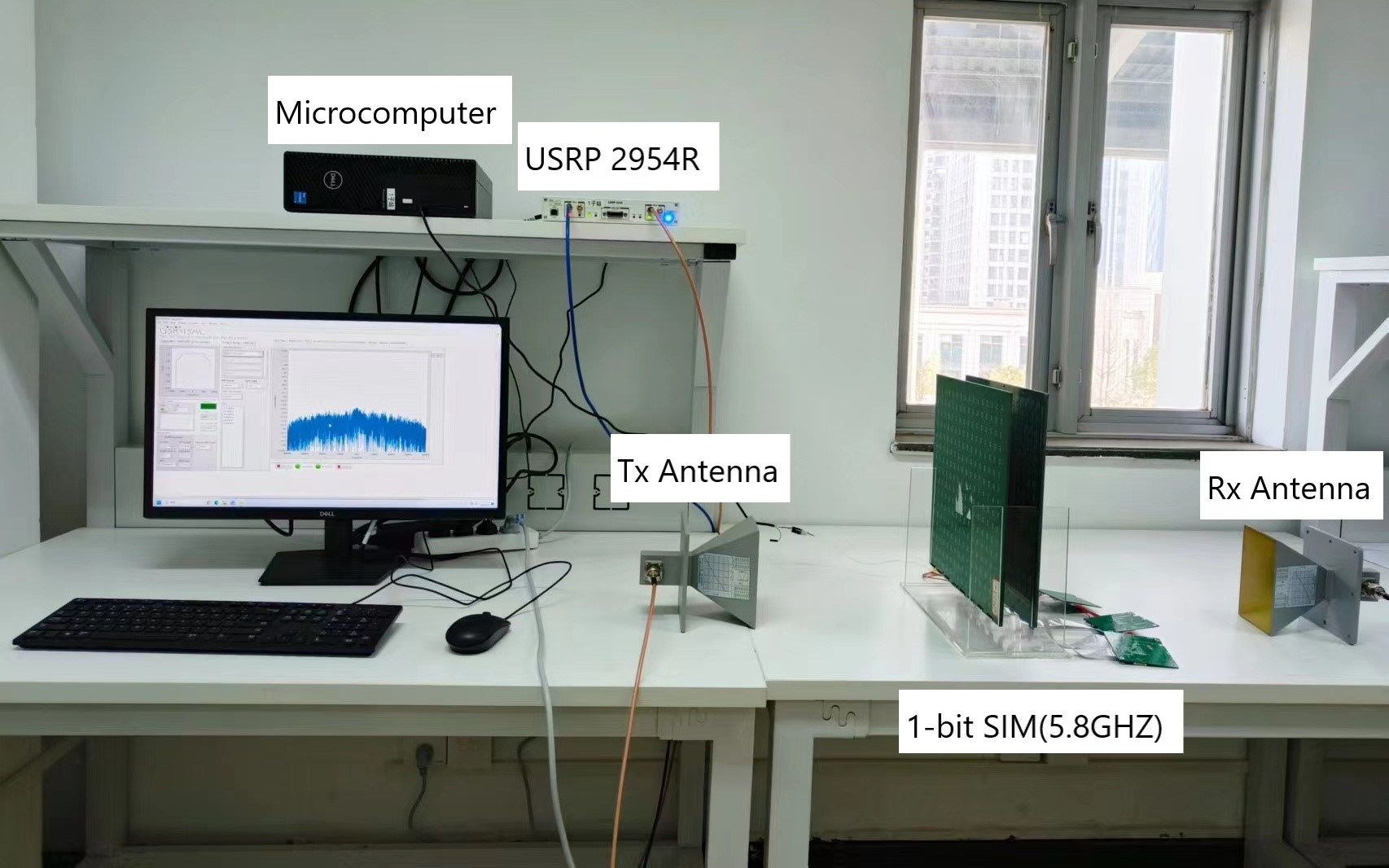}
\caption{The SIM-aided communication and sensing prototype system. (The RX serves as the object of sensing or the receiver for communication.)} 
\label{fig_SIM_exp_sys}
\vspace{-0.2 cm}
\end{figure}
% The dimensions of the transmissive RIS prototype measure $31 \times 31 cm^{2}$. 
% For the attainment of phase reconfigurability, individual control over each element within the array is essential. 
% This necessitates $196$ biasing lines, as each element requires a biasing line to apply DC voltages to two PIN diodes. 
% To manage these voltages independently, four logic circuit controlling boards are affixed to provide distinct DC voltages across $196$ channels. 
% These boards establish connections to the biasing lines through multiple connectors. 
% Stacking multiple $1$-bit transmissive RIS in Fig.~\ref{fig_SIM_struct}e together forms a $1$-bit SIM in Fig.~\ref{fig_SIM_struct}d.

The experiment system in Fig.~\ref{fig_SIM_exp_sys} uses a Universal Software Radio Peripheral(USRP) as the signal source, transmitting a signal at a frequency of $5.8$ GHz with a power of $10$ dBm. 
Each transmission coefficient matrix is derived through an exhaustive search in either communication or sensing experiments. 
Given the single-antenna system configuration, there is no beamforming scheme employed at the transmitter.

\subsection{Communication Experiment}
\begin{figure}[ht]
\vspace{-0.2 cm}
\centering
\includegraphics[width=2.1in]{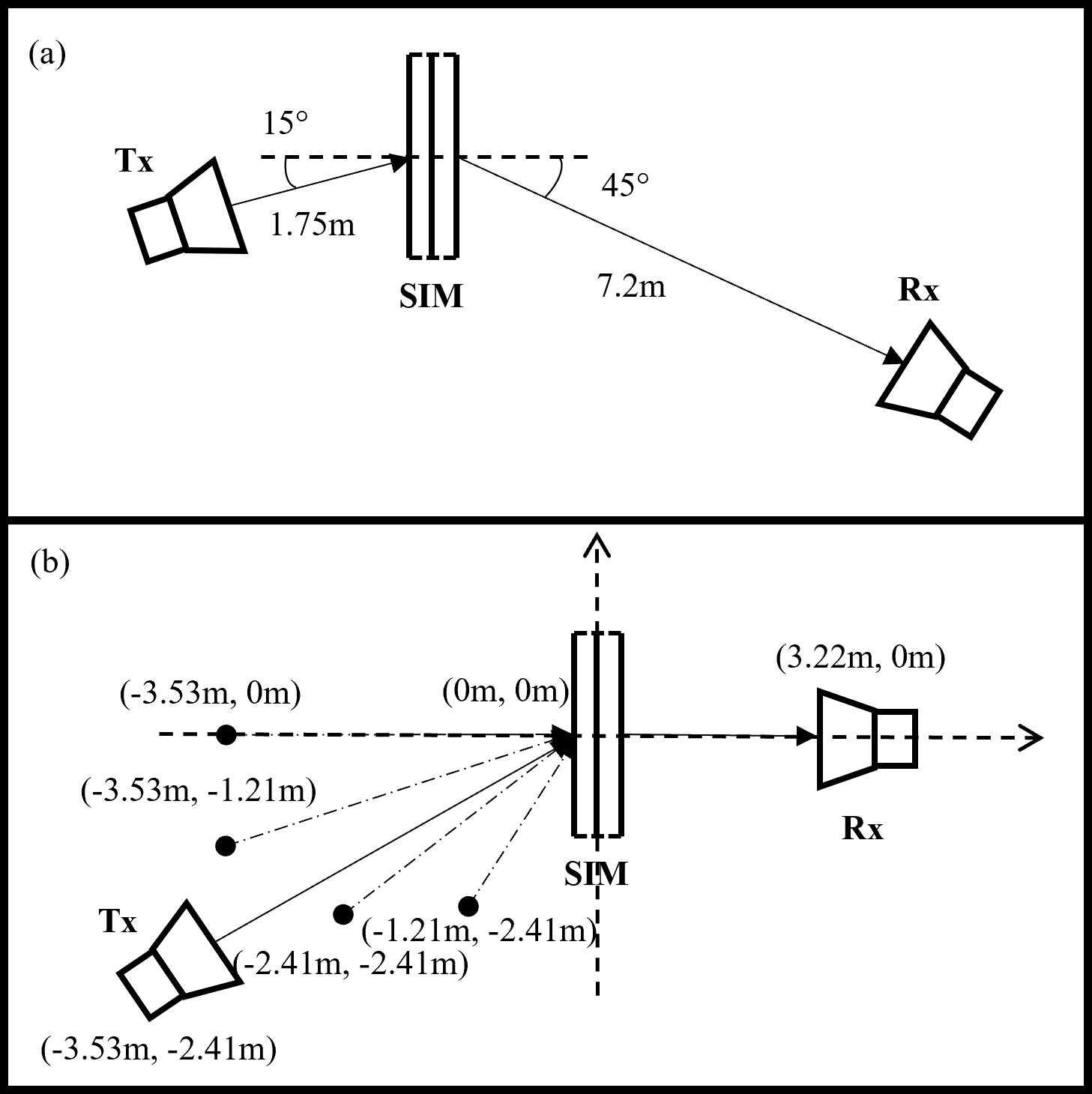}
\vspace{-0.2 cm}
\caption{Scenarios (a) Communication. (b) Sensing.} 
\label{fig_exp_com_sensing}
\vspace{-0.5 cm}
\end{figure}

The transmitted signal passes through a directional transmitter (TX) with an incident angle (relative to the normal) of $15^{\circ}$ onto a SIM board located at a distance of $1.75 {\rm m}$. 
From the last transmissive layer of the SIM, the signal exits at a $45^{\circ}$ angle (relative to the normal) and travels to the receiver (RX) located at a distance of $7.2 {\rm m}$. 
The USRP was used to process and analyze the received signal. The results are shown in Table~\ref{tab:table1} based on the experiment in Fig.~\ref{fig_exp_com_sensing}a.
% \begin{figure}[t]
% \centering
% \subfloat[]{\includegraphics[width=2.5in]{exp_com_tx}%
% \label{exp_tx}}
% \hspace{1in}\\
% \subfloat[]{\includegraphics[width=2.5in]{exp_com_rx}%
% \label{exp_rx}}
% \caption{System model of 1-bit SIM-Aided Communications. (a) TX-SIM. (b) SIM-RX.}
% \label{fig_SIM_com_exp}
% \end{figure}

\begin{table}[ht]

\caption{The performance of the SIM versus the number of layers and inter-layer spacing.\label{tab:table1}}
\centering
\begin{tabular}{cccccc}
\toprule
Num of layers & $\mathrm{\lambda}/4$ & $\mathrm{\lambda}/2$ &$\mathrm{\lambda}$ & $3\mathrm{\lambda}/2$ &$2\mathrm{\lambda}$\\
\midrule
1-layer &-53dBm &-53dBm &-53dBm &-53dBm &-53dBm\\
2-layer &-49dBm &-50dBm &-56.5dBm &-55dBm &-55dBm\\
3-layer &-46dBm &-49dBm &-59dBm &-58dBm &-57dBm\\
\bottomrule
\end{tabular}
\vspace{-0.2 cm}
\end{table}

\begin{figure}[ht]
\centering
\subfigure[]{\includegraphics[width=.49\linewidth]{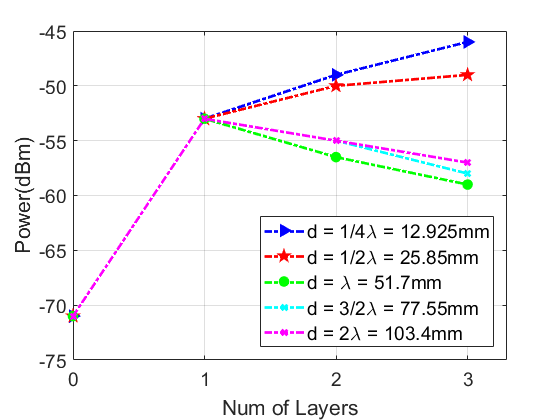}%
\label{exp_com_1}}
\subfigure[]{\includegraphics[width=.49\linewidth]{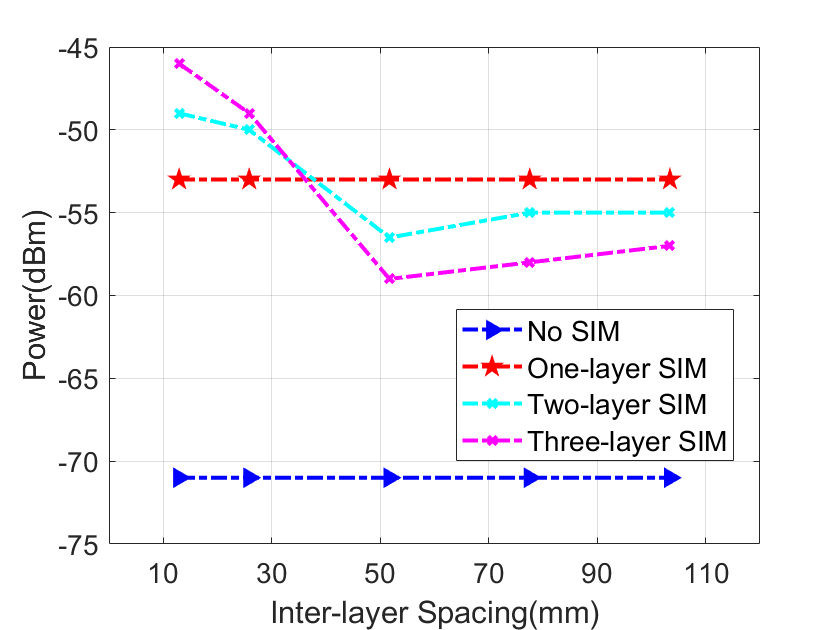}%
\label{exp_com_2}}
\vspace{-0.5 cm}
\caption{(a) Impact of the number of layers of the SIM on the communication signal. (b) Impact of the interlayer spacing of the SIM on the communication signal.}
\label{fig_SIM_com_exp_result}
\vspace{-0.6 cm}
\end{figure}
% The SIM introduces a discernible impedance in the experiment, as illustrated in Fig.~\ref{fig_SIM_com_exp_result}. 
It is worth noting that the gain depends on the inter-spacing between adjacent layers, as illustrated in Fig.~\ref{fig_SIM_com_exp_result}.
If the inter-spacing remains equal to or less than half the wavelength, the power at the RX increases with the number of layers. 
When the inter-spacing exceeds half of the wavelength, the trend of the curve is reversed. 
The rationale behind this trend is that the longer the inter-distance between the layers the higher the propagation losses between the layers, and the size of each layer of the SIM cannot compensate for the inter-layer attenuation in the considered experiments.
\vspace{-0.2 cm}
\subsection{Sensing Experiment}
\begin{table}[ht]
\vspace{-0.5 cm}
\caption{DoA estimation error for a different number of layers of the SIM.\label{tab:table2}}
\centering
\begin{tabular}{cccc}
\toprule
Coordinates of TX  & 1-layer & 2-layer & 3-layer \\
\midrule
(-3.53m, 0.00m) &$4.00^\circ$ &$3.00^\circ$ &$2.50^\circ$ \\
(-3.53m, -1.21m) &$5.72^\circ$ &$0.22^\circ$ &$1.22^\circ$ \\
(-3.53m, -2.41m) &$2.71^\circ$ &$1.79^\circ$ &$1.29^\circ$ \\
(-2.41m, -2.41m) &$3.50^\circ$ &$1.50^\circ$ &$1.50^\circ$ \\
(-1.22m, -2.41m) &$1.93^\circ$ &$0.93^\circ$ &$0.43^\circ$\\
\midrule
Average error  &$3.572^\circ$ &$1.488^\circ$ &$1.388^\circ$\\
\bottomrule
\end{tabular}
\vspace{-0.25 cm}
\end{table}

The system in Fig.~\ref{fig_SIM_exp_sys} is still utilized for the sensing experiment in Fig.~\ref{fig_exp_com_sensing}b. 
% and has the same parameters as the communication experiment. 
The experiment assumes fixed positions for both the SIM $(0 {\rm m}, 0 {\rm m})$ and the RX $(3.22 {\rm m}, 0 {\rm m})$. 
The inter-layer distance of the SIM is set to half a wavelength. 
Five groups of experimental data are collected by changing the position of the TX and the configuration of the SIM. 
The TX is regarded as a point target, which can be viewed as a special instance of an extended target.
The estimated DoA is then derived using the least squares estimator. 
The positions of the TX and the corresponding estimation errors are presented in Tab. \ref{tab:table2}. 
We observe that error for DoA estimation decreases with the number of transmissive layers of the SIM.

% \begin{figure}[t]
% \centering
% \subfloat[]{\includegraphics[width=2.5in]{exp_sensing_tx.jpg}%
% \label{exp_sensing_1}}
% \hspace{1in}\\
% \subfloat[]{\includegraphics[width=2.5in]{exp_sensing_rx.jpg}%
% \label{exp_sensing_2}}
% \caption{System model of 1-bit SIM-Aided Sensing. (a) TX-SIM. (b) SIM-RX.}
% \label{fig_SIM_sensing_exp}
% \end{figure}
\vspace{-0.1 cm}
\section{Conclusion}
This paper investigated an ISAC scenario with multiple communication users and an extended target. 
We proposed an optimization algorithm to jointly optimize the beamforming at the BS and the end-to-end transmission matrix of the SIM. Also, we have built a hardware platform for SIM and evaluated its performance through experiments.
Numerical and experimental results demonstrated that the increasing number of layers in the SIM can improve the ISAC performance.
\vspace{-0.5 cm}
% \bibitem{b1} G. Eason, B. Noble, and I. N. Sneddon, ``On certain integrals of Lipschitz-Hankel type involving products of Bessel functions,'' Phil. Trans. Roy. Soc. London, vol. A247, pp. 529--551, April 1955.
% \bibitem{b2} J. Clerk Maxwell, A Treatise on Electricity and Magnetism, 3rd ed., vol. 2. Oxford: Clarendon, 1892, pp.68--73.
% \bibitem{b3} I. S. Jacobs and C. P. Bean, ``Fine particles, thin films, and exchange anisotropy,'' in Magnetism, vol. III, G. T. Rado and H. Suhl, Eds. New York: Academic, 1963, pp. 271--350.
% \bibitem{b4} K. Elissa, ``Title of paper if known,'' unpublished.
% \bibitem{b5} R. Nicole, ``Title of paper with only first word capitalized,'' J. Name Stand. Abbrev., in press.
% \bibitem{b6} Y. Yorozu, M. Hirano, K. Oka, and Y. Tagawa, ``Electron spectroscopy studies on magneto-optical media and plastic substrate interface,'' IEEE Transl. J. Magn. Japan, vol. 2, pp. 740--741, August 1987 [Digests 9th Annual Conf. Magnetics Japan, p. 301, 1982].
% \bibitem{b7} M. Young, The Technical Writer's Handbook. Mill Valley, CA: University Science, 1989.
% \end{thebibliography}
\vspace{12pt}
% \color{red}
% IEEE conference templates contain guidance text for composing and formatting conference papers. Please ensure that all template text is removed from your conference paper prior to submission to the conference. Failure to remove the template text from your paper may result in your paper not being published.
\bibliographystyle{IEEEtran}
\bibliography{reference}

\end{document}